\documentclass[10pt]{article}
\usepackage{emlines2,bezier}
\title{\bf
Classical long-range interacting N-particle configurations and its
applications.}
\author{\it Sergey  S. Kokarev \\  Yaroslavl State Pedagogical University \\
\\Department of theoretical physics,  150000, Yaroslavl, Respublikanskaya 108,
r.409 \\ E-mail: sergey@yspu.yar.ru}
\date{}
\begin{document}
\maketitle

\begin{abstract}
We consider classical $N$-particle system with arbitrary  central pair
potential. Mechanical equilibrium condition in spherically-symmetric case leads
to a nonlinear integro-differential equation for concentration $n(r)$. For
special state equation  $p=an^{\gamma}(1+bn)$, original integro-differential
equation transforms into purely integral one. Its solution (under $b=0$) is
written through a row over interaction parameter. Physical conditions for its
convergency is discussed. For power-like potential kernal of integral operator
is calculated in apparent kind. The cases of Coulomb and harmonic potentials
are considered separately up to a third order. General scheme of application of
the theory for some astrophysical and cosmological problems is presented. Model
system with spherical potential pits is considered. By perturbation theory
first virial correction  $\delta n$ is calculated.
\end{abstract}

\section{Introduction}\label{intro}

Imperfect systems, whose interparticle (quasiparticle) interactions principally
determine its physical properties, are of great interest today. Their
distinctive peculiarities --- nonlinearity, nonadditivity of equilibrium
thermodynamic potentials and parameters, critical phenomenas, selfordering
effects etc -- endow this sort of system by nontrivial dynamics and interesting
observable effects. The last from the listed properties relates mainly to the
imperfect systems with {\it long-range} interactions, which can provide
orderization on scales about size of the system. Here and below the term
"long-range" will mean order of magnitude of interactions radius  equal or more
then size of a system. The principal difficulty of statistical investigation of
long-range interaction system (LRIS) concerns with a  very complicated kind of
statistical sum (integral) for long-range potentials. The well known virial
decomposition \cite{land1} become invalid for LRIS, because of relation of
interaction region to a volume of the system has order of unity. It was
spontaneously found, that methods, relevant to the problem mainly are related
to a different modifications of self-consistent field concepts. It is easily to
show, that this is, in some sense, unavoidable. As it has been mentioned
majority of customary properties of thermodynamic values are violated within
LRIS. Particularly, well known rule for calculation of pressure:
\begin{equation}\label{pressure}
p=-\left(\frac{\partial E}{\partial V}\right)_{S}
\end{equation}
becomes invalid or, more exactly, defines "generalized pressure". It is
expressed by the sum: $p=p^{(l)}+p^{(g)}$, where first term  is ordinary local
(contact) pressure, borned by classical or quantum motion of particles, while
second one is determined by a long range interaction of some picked volume with
environment. This decomposition corresponds to the subdividing of internal
energy $E=E^{(l)}+E^{(g)}$, where $E^{(l)}$ --- local internal energy,
$E^{(g)}$
--- interaction energy of a small picked subsystem with environment.
Now one can't understand  $p$ simply as a component of stress tensor of
continuum media, if stress tensor of long-range field is not also included into
consideration. The force components then will be $$
F_i=\oint\limits_{\Sigma=\partial
V}(\sigma^{(l)}_{ik}+t^{(f)}_{ik})\,d\Sigma^k=\int\limits_{V}{\rm
div}\,(\sigma^{(l)}+t^{(f)})_i\, dV,$$
  where $\sigma^{(l)}$  --- ordinary
continuum media stress tensor, $t^{(f)}$
--- field stress tensor, $V$ --- arbitrary volume inside the system.
Mechanical equilibrium condition reads:
\begin{equation}\label{compens}
{\rm div}(\sigma^{(l)}+t^{(f)})=0,
\end{equation}
that means intercompensation of local and global forces. But this resulting
force depends on some internal properties of the system, for example ---
density of "generalized charge" distribution and we unavoidably go to the
self-consistent problem.

Attempts of investigations of LRIS have been made in a different areas: for
dielectrics in electrostatic by thermodynamical methods in \cite{helm}, for
Newtonian gravity in balls stellar clusters by virial theorem \cite{einst}, by
variational methods with fields equation as constraints ---
\cite{nonid1,nonid2}, for quantum system  ---  \cite{hartri}-\cite{tel}.
General problems are both choice of interparticle interaction  or field
equation, and solving of self-consistent equations, which are in general case
nonlinear and integral.

In present paper we don't restrict ourself to any concrete physical system  and
consider abstract centrally-interacting structureless classical particles. It
is assumed, that pair potential satisfies some linear field equation, so that
superposition principle is valid.  We consider the case of identical particles,
with   pair potential:
\begin{equation}\label{pot}
\phi=\alpha g(r),
\end{equation}
where $\alpha$ --- generalized charge of a particle (the source of a long-range
field ), which is the same for all particles, $r$
--- distance between the particle-source and point of observation,
$g$ --- universal for all particles function, depending on the  distance and
some universal constant parameters. It is  postulated, that the system  after a
long time goes in equilibrium state, where all macroscopic parameters become
constant. We are interested by practically important case of macroscopic
spherically symmetry, when all macroscopic parameters (for example
concentration) depends only on a distance from a center of the symmetry.

Let in equilibrium state the system takes ball with radius $R$ and let $n(r)$
--- concentration of  particle, normalizing by the condition:
\begin{equation}\label{norm}
\int\limits_{V_{R}}n\,dV=4\pi\int\limits_{0}^{R}r^2n(r)\,dr=N,
\end{equation}
where $N\gg1$  --- full number of particles. In (\ref{norm}) central symmetry
is used.

Lets consider arbitrary point inside the system with radius-vector $\vec r$.
Potential $d\varphi$ at the point, producing by the particles inside some
remoted volume $dV'$ in neighborhood of some another point $\vec r'$ of the
system accordingly to  (\ref{pot}) is equal:
\begin{equation}\label{charge}
d\varphi(r)=g(|\vec r-\vec r'|)\,dL',
\end{equation}
where $dL'\equiv\alpha\cdot n(r')\cdot dV'$ --- total charge of $dV'$.
Superposition principle for the full potential at  $\vec r$ gives
\begin{equation}\label{super}
\varphi(r)=\int\limits_{V_{R}}g(|\vec r-\vec r'|)\,dL'.
\end{equation}
Taking into account apparent kind $|\vec r-\vec
r'|=\sqrt{r^2-2rr'\cos\theta+r'^2}$ and carring out angle integration, we get
\begin{equation}\label{pot1}
\varphi(r)=2\pi\alpha\int\limits_{0}^{R}G(r,r')n(r')r'^2\,dr'\equiv
2\pi\alpha\hat K[n]
\end{equation}
where
\begin{equation}\label{funcg}
G(r,r')=\int\limits_{-1}^{1}g(\sqrt{r^2-2rr'\xi+r'^2})\,d\xi=G(r',r)
\end{equation}
determines kernal  \footnote{Rigorously speaking kernal in (\ref{pot1}) is
$G(r,r')r'^2$, but for the brevity we'll understand  as kernal the function
$G$. Note, that $G(r',r)$ is, in fact, Green function for the linear equation,
which pair potential $\varphi$ obeys to. So by (\ref{pot}) we postulate special
Green function, that corresponds to a wide class of linear differential field
operators. } of integral operator $\hat K$.

 Attractive force, acting on a
volume element $dV$ from the effective field of the system is $$
dF=-\frac{\partial\varphi}{\partial r}dL=-\frac{\partial\varphi}{\partial r}
\cdot \alpha\cdot n(r)\cdot dV, $$ and its volume density
\begin{equation}\label{force}
f=\frac{dF}{dV}=-\frac{\partial\varphi}{\partial r}\alpha\cdot n(r)=
-2\pi\alpha^2n(r)\int\limits_{0}^{R} \frac{\partial G(r,r')}{\partial
r}r'^2n(r')\, dr'
\end{equation}
Equation (\ref{compens}) of mechanical equilibrium, which in our case is
reduced to $-dp/dr+f=0$, with local\footnote{Here and below $p$ will mean
ordinary local pressure.} pressure $p$, takes the form:
\begin{equation}\label{eq1}
\frac{dp}{dr}+2\pi\alpha^2 n(r)\int\limits_{0}^{R} \frac{\partial
G(r,r')}{\partial r}r'^2n(r')\, dr'=0
\end{equation}
The obtained equation belongs to a class of integro-differential ones. Really,
pressure in general case is connected with concentration by thermal state
equation $p=p(n)$, which should be accounted by statistical methods. In our
case it is determined also by long-range potential (its derivatives), whose
influence on state equation is very complicated.

\section{Simplifying of equilibrium equation}\label{part}

To make analytical investigation of eq.(\ref{eq1}) some simplifying assumptions
are necessary:

1) Assume the local field has no influence on local state equation. This
assumption will be valid in case of not strong fields and its gradients. In
fact our consideration concerns with zero-th term of Taylor row of $p$ over
$\varphi$ (or, more strictly its derivatives).

2) Lets restrict our attention to a  wide class of state equations of the form:
\begin{equation}\label{eqst}
p=an^{\gamma}(1+bn)
\end{equation}
where $a$, $b$ and  $\gamma$ --- phenomenological constants. Note, that $a$ has
a sense of temperature multiplier, which constantness within the system
automatically implies its heat equilibrium. Expression (\ref{eqst}) should be
understood as the first two terms of effective virial expansion. For a not
large concentrations second term much less then first one and so it influence
on equilibrium distribution can be accounted by perturbations methods, with
background, corresponding to the state equation $p=an^{\gamma}$.

Stress, that thermal state equation (\ref{eqst}) is postulated by us, but not
calculated as it should be done in fundamental theory. So our approach should
be called {\it semi-self-consistent}: we take into account influence of
long-range potential on concentration, but not on state equation.

3) Lets take function  $g(r)$ from a ring of a generalized polynoms:
\begin{equation}\label{gpot}
g(r)=\sum\limits_{i}^{}\varepsilon_{i}r^{\lambda_{i}},
\end{equation}
where $\epsilon_{i},\ \lambda_i$ --- arbitrary, but similar for all particles
real parameters of potential.

The first and second assumptions (under $b=0$) in (\ref{eqst}) lead to
equation: $$ a\gamma n^{\gamma-1}n'+2\pi\alpha^2 n(r)\int\limits_{0}^{R}
\frac{\partial G(r,r')}{\partial r}r'^2n(r')\, dr'=0 $$ or,  after integration
over  $r$:
\begin{equation}\label{gneq1}
n^{\gamma-1}=-s\hat K[n]+A
\end{equation}
in case $\gamma\neq1$, and
\begin{equation}\label{geq1}
n=\bar{A}e^{-\bar s\hat K[n]},
\end{equation}
in case $\gamma=1$. Here $A$ and $\bar A$ --- yet undefined integration
constants, $s=2\pi\alpha^2(\gamma-1)/a\gamma$,
\begin{equation}\label{bs}\bar s=2\pi\alpha^2/a.
\end{equation}
It easily to see, that equation (\ref{geq1}) is Boltzmann distribution written
in self-consistent form. Really identifying  (\ref{geq1}) with  $n=n_0{\rm
exp}(-U/kT)$  under $\bar A=n_0$,
\begin{equation}\label{a}
a=kT,
\end{equation}
$U=2\pi\alpha^2\hat K[n]$ (see (\ref{pot1}), (\ref{bs})), we obtain that this
is true.

The third simplifying assumption lets us by elementary integration calculate
$G$. By linearity $G$ over $g$ it is sufficiently to carry out all calculations
for one arbitrary power $\lambda$ in (\ref{gpot}). The result is:
\begin{equation}\label{kern}
G(r,r')= \int\limits_{-1}^{1}\varepsilon(r^2-2rr'\xi+r'^2)^{\lambda/2}\,d\xi=
\left\{
\begin{array}{lr}
\displaystyle{\frac{\varepsilon}{(\lambda+2)rr'}[(r+r')^{\lambda+2}-|r-r'|^{\lambda+2}]},&
\lambda\neq-2;\\
\displaystyle{\frac{\varepsilon}{rr'}\ln\left|\frac{r+r'}{r-r'}\right|},&
\lambda=-2
\end{array}
\right.
\end{equation}

\section{General properties of equilibrium equations}

In this section we present some general consequences of (\ref{gneq1}) and
(\ref{geq1}), which are independent of any assumptions concerning with
perturbation theory (see following section). More exactly we are interested by
possibility of {\it cusps} and {\it finite radius} $R$ of the system, that is
actual in galaxy and stellar dynamics \cite{og,din1,din2,din3}. The necessary
conditions can be formulated by the two following simple theorems:
\vspace{0.4cm}

{\bf Theorem 1.} {\it If $g(r)$ is bounded from the up, then for $\gamma\le1$
configuration with finite $R$ is impossible.} \vspace{0.4cm}

{\bf Theorem 2.} {\it If $g(r)$ is bounded from the down, then for $\gamma\ge1$
configuration with cusp is impossible.}\vspace{0.4cm}

{\bf Proof.} Condition for finite $R$ is $n(R)=0\ (R<\infty).$ From
(\ref{gneq1}) and (\ref{geq1}) under $\gamma\le1$ it is followed, that this
possible if (and only if)  $\left.\hat K[n]\right|_{r=R}\rightarrow+\infty.$
Bounded from the up potential $g$ can be made negative on its whole domain of
definition by finite shift: it only will redetermine integration constants $A$
and $\bar A$. Such negative $g$ will give negative $G$  and negative  $\hat
K[n],$ since $n(r)\ge0.$ So, $\hat K[n],$ even it will have singular points
will negative there: $\left.\hat K[n]\right|_{r=R}\rightarrow-\infty,$ that
means impossibility of finite radius.

By the similar manner one can proof that under $\gamma\ge1$ and $g$ bounded
from the down $\hat K[n]$ at best gives: $\left.\hat
K[n]\right|_{r=0}\rightarrow+\infty,$ that is opposite to that  is required by
cusps condition $\left.\hat K[n]\right|_{r=0}\rightarrow-\infty.$

The important consequence from the theorems is: {\it finite configuration with
cusp in case $\gamma=1$ is possible, when $g$ is unbounded both from the up,
and from the down.} This fact suggests, that the so called {\it force
softening} procedure \cite{soft}, introducing in N-body simulation scheme to
avoid local force infinities  under collisions and to short computer time, can
(under certain circumstances) drastically change global properties of
investigated system.

\section{General form of solutions}\label{vdv}

Let build at first a general solution for the case (\ref{geq1}). For this
purpose we write it through the formal row over powers of interaction
parameter\footnote{Inspite of dimension character of $\bar s$ we prefer it to
dimensionless ordering parameter $\xi=\alpha^2R^{\lambda}/a$ (see below), since
$R$ can be infinite.} $\bar s$:
\begin{equation}\label{row}
n=X_{0}+X_{1}\bar s+X_{2}\bar s^2+\dots,
\end{equation}
where $X_i$ --- yet unknown functions $r$. Expand then the exponent on $\hat K$
in (\ref{geq1}) in the Taylors row:
\begin{equation}\label{exp}
e^{-\bar s\hat K[n]}=1-\frac{\bar s\hat K[n]}{1!}+\frac{\bar s^2\hat
K^2[n]}{2!}+\dots=
\end{equation}
$$ 1-\frac{\bar s(\hat K[X_0]+\bar sK[X_1]+\dots)}{1!}+ \frac{\bar s^2(\hat
K[X_0]+\bar s\hat K[X_1]+\dots)^2}{2!}-\dots $$ To equate coefficients from the
left and right in (\ref{geq1}) after substituting (\ref{row}) and (\ref{exp}),
it is necessary to reexpand every term of (\ref{exp}) over $\bar s$. The
following useful formula, which can be proved by mathematical induction, is
convenient:
\begin{equation}\label{formula}
(x_0+x_1\bar s+x_2\bar s^2+\dots)^p=x_0^p+
\end{equation}
$$ \sum\limits_{n=1}^{\infty} \frac{\bar s^n}{n!}\sum\limits_{(\vec k,\vec
n)=n} \frac{p!(1!)^{k_1}(2!)^{k_2}\cdots(n!)^{k_n}}{(p-k_1-\dots-k_n)!}
B^{n}_{k_1k_2\dots k_n}x_0^{p-k_1-\dots-k_n}x_1^{k_1}\cdots x_n^{k_n}. $$ In
this expression $\vec k\equiv(k_1,\dots,k_n)$, $\vec n\equiv(1,\dots,n)$
--- integer-values $n$-dimensional vector, $(\vec a,\vec b)$ --- euclidian scalar
product,  coefficients $B^n_{k_1k_2\dots k_n}$ can be calculated with the help
of recurrent formulae, given in Appendix 1. For an integer positive $p$
summation is carried out only for these $\vec k$, which satisfies the
nonequality $p-k_1-\dots-k_n\ge0$. It is easily to check that for particular
cases $p\in{\cal N}$ and $p=-1$ under $x_i=0$ for all $i=2,\dots$
(\ref{formula}) reproduces well known Newton binom and formula for infinite
geometrical progression correspondingly.

Substituting (\ref{row}) and (\ref{exp}) into (\ref{geq1}), taking into account
(\ref{formula}) and equating coefficients for equal powers $\bar s$ in
(\ref{geq1}) from the left and right, we get  the system  of recurrent
relations on unknown coefficients in (\ref{row}):
\[
X_0=\bar A;\ \ X_1=-\bar A\hat K[X_0];
\]
\begin{equation}\label{gsol1}
\ \ X_j=\bar A\left[\sum\limits_{p=1}^{j-1} \frac{(-1)^{p}}{(j-p)!}
\sum\limits_{(\vec k,\vec{j-p})=j-p}
\frac{(1!)^{k_1}(2!)^{k_2}\cdots((j-p)!)^{k_{j-p}}}{(p-k_1-\dots-k_{j-p})!}
B^{j-p}_{k_1\dots k_{j-p}}\times\right.
\end{equation}
\[ \left.(\hat K[X_0])^{p-k_1-\dots-k_{j-p}}(\hat K[X_1])^{k_1}\cdots(\hat
K[X_{j-p}])^{k_{j-p}}+ \frac{(-1)^{j}}{j!}(\hat K[X_0])^j\right]. \] For the
first four equations of the infinite chain (\ref{gsol1}) gives:
\begin{equation}\label{ex1}
\begin{array}{l}
X_0=\bar A;\\
\\
X_1=\bar A \hat K[X_0]=-\bar A^2K;\\
\\
X_2=\bar A(-\hat K[X_1]+\frac{1}{2}\hat K^2[X_0])=\bar
A^3\left[K_2+\frac{1}{2}K^2\right];\\
\\
X_3=\bar A(-\hat K[X_2]+\hat K[X_0]\cdot \hat K[X_1]-\frac{1}{3!}\hat
K^3[X_0]))=\\
\\
-\bar A^4\left[K_3+\frac{1}{2}K[K^2]+K\cdot K_2+\frac{1}{3!}K^3\right]\\ \cdots
\end{array}
\end{equation}
Here $ K\equiv \hat K[1];\ \ K_n\equiv\hat K^n[1].$

The same way  is appropriate for the more general case  (\ref{gneq1}) of
equilibrium equation. Substitution of decomposition (\ref{row}) (we omit a bar
over letters)   and equation of coefficients for all equal powers of $s$ with
(\ref{formula}) gives:
\[
X_0^{\gamma-1}=A;\
\]
\[
\frac{1}{n!}\sum\limits_{(\vec k,\vec n)=n}
\frac{(\gamma-1)!(1!)^{k_1}(2!)^{k_2}\cdots(n!)^{k_n}}
{(\gamma-1-k_1-\cdots-k_n)!} B^n_{k_1\dots
k_n}X_0^{\gamma-1-k_1-\cdots-k_n}X_1^{k_1}\cdots X_n^{k_n}=-K[X_{n-1}],
\] or in resolved form:
\begin{equation}\label{gsol2}
X_n=-\frac{X_0^{2-\gamma}}{(\gamma-1)}\left(\hat
K[X_{n-1}]+\frac{1}{n!}\sum\limits_{(\vec
k,\vec{n-1})=n}^{}\frac{(\gamma-1)!(1!)^{k_1}\cdots((n-1)!)^{k_{n-1}}}{(\gamma-1-\cdots
-k_{n-1})!}B^{n}_{k_1\cdots k_{n-1}0}\times\right.
\end{equation}
\[
\left.X_0^{\gamma-1-\cdots-k_{n-1}}X_{1}^{k_1}\cdots X^{k_{n-1}}_{n-1}\right).
\] All noninteger factorials in (\ref{gsol2}) should be understood as infinite
products $a!\equiv a(a-1)(a-2)\cdots$. The first four equation of the infinite
chain has the form:
\begin{equation}\label{ex2}
\begin{array}{l}
X_0=A^{\frac{1}{\gamma-1}}\equiv\tilde X_0;\\
\\
X_1=-\frac{\displaystyle\tilde X_0^{3-\gamma}}{\displaystyle\gamma-1}K;\\
\\
X_2=\frac{\displaystyle\tilde
X_0^{5-2\gamma}}{\displaystyle(\gamma-1)^2}\left[K_2-\frac{\displaystyle
1}{\displaystyle 2}(\displaystyle \gamma-2)K^2\right];\\
\\
X_3=-\frac{\displaystyle\tilde
X_0^{7-3\gamma}}{\displaystyle(\gamma-1)^3}\left[K_3-\frac{\displaystyle
1}{\displaystyle 2}(\displaystyle \gamma-2) \hat
K[K^2]+\frac{\displaystyle(\gamma-2)(2\gamma-3)}{\displaystyle6}K^3-(\gamma-2)K\cdot
K_2\right].
\end{array}
\end{equation}

At the end of the paragraph lets discuss some general physical considerations,
relating to the convergency question of the row (\ref{row})\footnote{Note, that
direct mathematical investigation of the rows (\ref{gsol1})-(\ref{gsol2})
convergency is problematic due to a very complicated structure of its
coefficients. Even the evaluation of a number of nonzero $B^n$ for arbitrary
fixed $n$, presented in Appendix 1, requires rather long calculations on
n-dimensional integer-valued lattice.}. If generalized charges of a system will
be too large or potential will be too rapidly fall in neighborhood of a
particle (see following paragraph), then the configuration can not be
equilibrium: there is singular condensate of the particles (with zero classical
entropy) and row (\ref{row}) is diverging. As a characteristic parameter, which
is responsible for the
 such condensate  forming, the value
\[\xi=\frac{\alpha^2R^{\lambda}}{a}\sim\frac{u_{R}}{kT},\] should  be taken, where $u_R$ ---
potential energy of a pair particles at a distance of size of the system. The
parameter $\xi$ is also characterize the influence of a long-range ordering
forces, in comparison with influence of disordering heat chaotic particles
motion inside the system. So, $\xi$ is natural to be called  {\it ordering
parameter.}

On the other hand, an energy of long-range interaction of a particle with the
whole system must be sufficiently large to provide bound state of the
particles. In opposite case the system will be diffused. Corresponding
parameters is
\[
\Xi=\frac{U_R}{a}\sim\frac{U_R}{kT},
\]
where $U_R$ --- potential energy of interaction of a particle with a whole
system. So reasonable physical conditions of equilibrium (and convergency of
the rows) are:
\[
\xi\ll 1\ll \Xi.
\]

\section{The cases $\lambda=-1$ and $\lambda=2$} \label{cul}

\begin{figure}[htb]
\centering \unitlength=1.00mm \special{em:linewidth 0.4pt}
\linethickness{0.4pt}
\begin{picture}(109.00,42.00)
\put(36.67,3.33){\makebox(0,0)[ct]{$R$}}
\bezier{176}(12.33,32.00)(29.67,32.67)(36.33,7.33)
\bezier{76}(68.33,30.67)(75.67,31.67)(82.33,22.33)
\bezier{88}(94.33,7.33)(90.00,6.67)(82.67,22.33)
\put(95.00,4.00){\makebox(0,0)[ct]{$R$}}
\emline{12.00}{42.00}{1}{12.00}{7.00}{2}
\emline{12.00}{7.00}{3}{49.00}{7.00}{4} \emline{49.00}{7.00}{5}{45.00}{9.00}{6}
\emline{49.00}{7.00}{7}{45.00}{5.00}{8}
\emline{12.00}{42.00}{9}{10.00}{37.00}{10}
\emline{12.00}{42.00}{11}{14.00}{37.00}{12}
\put(16.00,39.00){\makebox(0,0)[lc]{$n(r)$}}
\put(47.00,12.00){\makebox(0,0)[cb]{$r$}}
\emline{68.00}{42.00}{13}{68.00}{7.00}{14}
\emline{68.00}{7.00}{15}{109.00}{7.00}{16}
\emline{109.00}{7.00}{17}{105.00}{9.00}{18}
\emline{109.00}{7.00}{19}{105.00}{5.00}{20}
\emline{68.00}{42.00}{21}{70.00}{38.00}{22}
\emline{68.00}{42.00}{23}{66.00}{38.00}{24}
\put(73.00,40.00){\makebox(0,0)[cc]{$L$}}
\put(107.00,12.00){\makebox(0,0)[cb]{$r$}}
\put(31.00,39.00){\makebox(0,0)[cc]{(a)}}
\put(88.00,40.00){\makebox(0,0)[cc]{(b)}}
\end{picture}
\caption{\scriptsize (а) --- schematic dependency of $n(r)$ for interparticle
Coulomb and harmonic potentials in the first approximation; (b)
--- theoretical radial profile of luminosity for galaxies $E0$ for the same potentials.}\label{graf}
\end{figure}
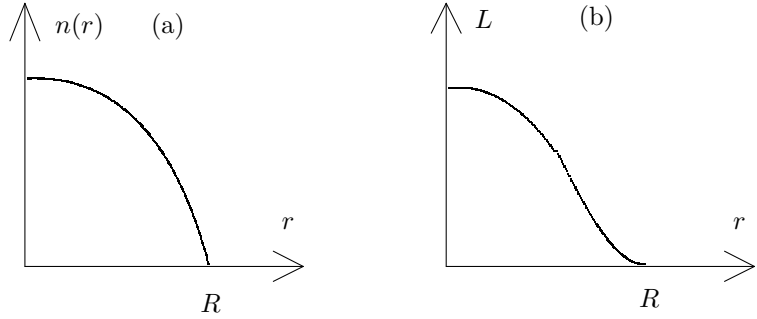

Due to simplicity and practical importance particular interest  have particles
with Coulomb and harmonic potentials. Lets calculate the first nontrivial
approximation $X_1$ for $\gamma=1$. As it is seen from (\ref{ex1}) the matter
is reduced to a calculation of $K$. Simple analysis shows, that $K\to\infty$
under $\lambda\le-3$. In this case all terms  of the formal row (\ref{row})
become infinite, that expresses nonstability of configuration with these
potentials: all particles falls onto the center. In case  $\lambda>-3$
elementary integration gives:
\begin{equation}\label{k}
K=\left\{
\begin{array}{lr}
\displaystyle \frac{\varepsilon}{(\lambda+2)r}\left\{\frac{1}{\lambda+4}
((R+r)^{\lambda+4}-(R-r)^{\lambda+4})-\frac{r}{\lambda+3}((R+r)^{\lambda+3}+
(R-r)^{\lambda+3})\right\},&\\ &\\ \lambda\neq-2;&\\ &\\ \frac{\displaystyle
1}{\displaystyle 2r}\left[(R^2-r^2)\ln\frac{\displaystyle R+r}{\displaystyle
R-r}+2Rr\right],&\ \\ &\\ \lambda=-2.
\end{array}
\right.
\end{equation}
For  $\lambda=-1$ we have $K=\varepsilon R^2(1-z^2/3)$. For $\lambda=2$,
$K=\varepsilon(2R^5/5)(1+5z^2/3)$, $z=r/R$. In the first case one should put
$\varepsilon=-1$, in second $\varepsilon=1$ to provide attractive character of
interaction. For the former accordingly to (\ref{row}) and (\ref{ex1})
expression for $n(r)$ takes the form:
\begin{equation}\label{approx1}
n\approx n_0-n_0^2\bar sK =n_0+n_0^2\bar sR^2\left(1-\frac{z^2}{3}\right),
\end{equation}
which must vanish at $z=1$. It gives $n_0=-3/2R^2\bar s$. Substituting in
(\ref{approx1}), we get
\begin{equation}\label{conc1}
n_{\rm c}=\frac{3}{4R^2\bar s}\left(1-z^2\right).
\end{equation}
Writing normalizing condition (\ref{norm}) for  (\ref{conc1}) we get the
following first approximation for relation between equilibrium size of a system
and number of particles:
\begin{equation}\label{size1}
R_{\rm c}=\frac{5\bar s}{2\pi}N
\end{equation}
Similarly for the case of harmonic pair potential we obtain:
\begin{equation}\label{conc2}
n_{\rm h}=\frac{75}{128\bar s R^5}\left(1-z^2\right)
\end{equation}
and
\begin{equation}\label{size2}
R_{\rm h}=\sqrt{\frac{5\pi}{16\bar sN}}.
\end{equation}
Schematic curve for  the both cases is shown in fig.\ref{graf}(a).

The obtained expressions for equilibrium sizes of a system shows its strong
nonadditivity, that has been mentioned in introduction. It is interesting, that
equilibrium size of the system of bounded harmonic oscillators  is decrease,
when N increase, as it follows from (\ref{size2}).

Note, that within central two-body problem there is interesting symmetry
between this types of potentials: trajectories can be transformed to each other
by the so called Bohlin's transformation \cite{ar}. As it can be seen from
(\ref{conc1}) and (\ref{conc2}) the similar symmetry is revealed also within
N-body problem at least in first approximation. As it follows from the theorems
of the previous section, Coulomb system has $R=\infty$ and finite $R$ in
(\ref{size1}) is induced by approximation, while harmonic system has'nt cusp at
the center.  In fact, it  means that Coulomb system  of $N<\infty$ attractive
particles totally diffuse in equilibrium. It theoretically support the fact,
that galaxy and stars formations are looked as essentially nonequilibrium
systems. Futher we'll use expressions (\ref{conc1})-(\ref{conc2}) only to
demonstrate the method, without any aims to reproduce observations.

Using integrals $K_n$ from Appendix2, we can calculate the following
approximation for $n$. The third\footnote{The second one is innormalizable on
finite size R.} has the form $$ n\approx\frac{11}{R^2\bar
s}\left(1-1.26z^2+0.29 z^4-0.03z^6\right);\ \ R\approx\frac{\bar s N}{5.2\pi}
$$ and for harmonic potential: $$ n\approx\frac{0.52}{\bar
sR^5}\left(1-0.95z^2-0.14z^4-0.05z^6\right);\ \
R\approx\sqrt{\frac{0.24\pi}{\bar sN}}. $$

\section{Application in cosmology}\label{cosm}

Our model may be relevant to the systems of a large ($N\gg1$) classical
interacting objects, without any reference to its real scale. As an example,
let consider cosmological application of the theory
--- $E0$ galaxies or ball clusters.
Statistical approach to such systems has been applied  in  \cite{og}. Rough
consideration, based on virial theorem was made by Einstein in \cite{einst}.

For a comparing the theory with observable  datas it is necessary to translate
it in terms of visual optical values, for example {\it absolute surface
luminosity.} For a model, where all stars have the same overage characteristics
it is easily to obtain the following formula, relating surface luminosity  with
concentration:
\begin{equation}\label{blesk}
L=l_0\int\limits_{-\sqrt{R^2-\rho^2}}^{\sqrt{R^2-\rho^2}}n(\sqrt{\rho^2+\xi^2})\,
d\xi,
\end{equation}
where $l_0$ --- absolute luminosity of a single star, $\rho$
--- polar radius in the coordinate frame, related to the picture plane.

Substituting  of (\ref{conc1}) or   (\ref{conc2}) in  (\ref{blesk}) leads to
the following expression:
\begin{equation}\label{form}
L=A\left(1-\zeta^2\right)^{3/2},  \ \ \zeta=\frac{\rho}{R},
\end{equation}
where $$ A_{\rm coul}=\frac{l_0}{Rs};\ \ A_{\rm harm}=\frac{25 l_0}{32R^4s}, $$
which should be compared with experimental functions $L(\rho)$. In
fig.\ref{graf}(b) the curve (\ref{form}) is schematically shown. This curve is
well applied for a central part of cluster, where  number  of stars is
sufficient to form self-consistent field, but individual stars are well
resolved. Boundary region of a cluster is turn off from our consideration et
al, since there are large fluctuations. The problem of separation and
description of clusters  core and bondary is considered in details in
\cite{og}.

The method is also applicable for an investigation of a dark matter problem
\cite{obsor}, which had been mentioned yet in Einstein cited work \cite{einst}.
The problem is inverse with respect to preceding one: using photometric datas
one should reconstruct pair potential of interaction of astrophysical objects.
Below we formulate general rough scheme of the approach, defering details and
corrections for a future investigations.

As it is well known from rotational curves analysis and from another
astrophysical datas luminous mass of a galaxy presents it smaller part, while
remained  one (about 90\%) is confined in a dark halo, which extends  to a
distances about some radiuses of the visual central part. We assume, that
equilibrium properties of a dark matter --- its distribution and equilibrium
size are practically independent from the luminous ones, so that the last can
be viewed as probe subsystem at the dark background. Lets dark matter particles
interact with each other by means of potential $\tilde\phi$, where "tilde" will
denote values, related to dark matter. Then using general formulas of our
approach, one can calculate dark matter distribution $\tilde n$. This last
create at every point additional potential for luminous matter, which is self
interacting with Coulomb potential. The self-consistent "dark  potential"
$\tilde\varphi$, should be taken into account in  expression (\ref{pot1}) for a
total potential:
\begin{equation}\label{potsum}
\varphi(r)=2\pi\alpha\int\limits_{0}^{R}G(r,r')n(r')r'^2\,dr'
+\tilde\varphi\equiv 2\pi\alpha\hat K[n] +\tilde\varphi.
\end{equation}
The potential $\tilde\varphi$ when $\tilde n$ is known can be calculated with
the help of (\ref{pot1}), with interaction and temperature  parameters
$\tilde\alpha,\,\tilde a$ and kern $\tilde G$, taken in general form
(\ref{kern}) with parameters $\tilde\lambda$ and $\tilde\varepsilon$. These
last are, in fact, to be found. For the above considered Coulomb and harmonic
potentials simple calculations give:
\begin{equation}\label{darkpot}
\tilde\varphi_{\rm coul}=-\frac{3\tilde a}{4\tilde\alpha} \left(\frac{\tilde
z^4}{10}-\frac{\tilde z^2}{3}+\frac{1}{2}\right)+\mbox{\rm const};\ \
\tilde\varphi_{\rm harm}=\frac{15\tilde a}{224\tilde\alpha}\left(
1+\frac{7}{3}\tilde z^2\right)+\mbox{\rm const},
\end{equation}
where  $\tilde z=r/\tilde R$, $\tilde R$ --- dark halo radius. Integration of
equilibrium equation for luminous matter with potential (\ref{potsum}) in case
 $\gamma=1$ leads to the equation, generalizing (\ref{geq1}) (we omit bar over letters):
\begin{equation}\label{ggeq1}
n=Ae^{-s\hat K[n]-\kappa\tilde\varphi},
\end{equation}
where $\kappa=\alpha'/a$, $\alpha'$ --- new interaction constant of luminous
and dark matters. Introducing new function $n^*=ne^{\kappa\tilde\varphi}$, and
new kernal $G^*(r,r')=G(r,r')e^{-\kappa\tilde\varphi(r)}$, one can put
(\ref{ggeq1}) to the form (\ref{geq1}):
\begin{equation}\label{ggeq1*}
n^*=Ae^{-s\hat K^*[n^*]}.
\end{equation}
If luminous particles are interact with each other through Coulomb potential
then with equality $K^*=e^{-\kappa\tilde\varphi}K$, first approximation over
$s$ gives:
\begin{equation}\label{concdark}
n=F_2(\tilde\varphi)\frac{3}{4R^2s}\left(1-\frac{r^2}{R^2}F_1(\tilde\varphi)\right)
\end{equation}
--- expression for luminous matter distribution with correction formfactors.
These last are
$F_1(\tilde\varphi)=(3-2e^{\kappa\tilde\varphi})^{-1}$,
$F_2=e^{-2\kappa\tilde\varphi}/F_1=3e^{-2\kappa\tilde\varphi}-2e^{-\kappa\tilde\varphi}$,
where norming of potential
is $\tilde\varphi(R)=0$. Under $\kappa=0$ (interaction between luminous and dark matter is
absent) (\ref{concdark}) transforms into (\ref{conc1}). Then one should
substitute expression (\ref{concdark}) into (\ref{blesk}) for surface luminousity
calculation. The result can be represented in the following convenient for comparison
form:
$$ L(\zeta)=L_0(\zeta)+\Delta L(\zeta), $$
where correction
\begin{equation}\label{blesk1}
\Delta L=-\frac{3l_0}{4Rs} \int\limits_{-\sqrt{1-\zeta^2}}^{\sqrt{1-\zeta^2}}
\left(1+z^2(F_1F_2-1)-F_2\right)\,d\xi|_{z=\sqrt{\zeta^2+\xi^2}},
\end{equation}
and $L_0$ --- surface luminosity (\ref{form}) in absence of dark matter. For
example, we present expression for $\Delta L$, calculated on a dark matter
background with Coulomb and harmonic potentials. Under the approximations
$R/\tilde R=\varepsilon\ll1$, which very simplifies final expression we have:
\begin{equation}\label{Lgarm}
\Delta L=\bar
A\cdot\frac{l_0\nu\varepsilon^2}{Rs}(1-\zeta^2)^{3/2}\left(1-\frac{4}{9}\zeta^2\right);\
\
\bar A_{\rm coul}= \frac{9}{10};\ \ \bar A_{\rm harm}=\frac{9}{16}.
\end{equation}

where $\nu=(\alpha \tilde a/\tilde \alpha a)$. Comparison (\ref{Lgarm}) with
observations gives us  possibility for estimations of some combination of
interaction parameters and kind of potential. Note, that proposed approach can
can be complement to the so called MOND (Modified Newton Dynamics), where one
deals with  modified Newtonian pair potential too \cite{milgrom}. Obviously,
the presented scheme is too rough and demands its following improving: account
of a back reaction of a luminous matter on a dark one, more wide class of model
potential, introducing of multicomponent systems etc..

\section{The model with potential pits.}\label{nonid}

\begin{figure}[htb]
\centering \unitlength=1.00mm \special{em:linewidth 0.4pt}
\linethickness{0.4pt}
\begin{picture}(105.33,44.33)
\emline{5.33}{5.67}{1}{5.33}{44.00}{2} \emline{5.33}{44.00}{3}{3.33}{38.33}{4}
\emline{5.33}{44.00}{5}{7.33}{38.67}{6} \emline{5.33}{6.00}{7}{45.67}{6.00}{8}
\emline{45.67}{6.00}{9}{41.00}{7.67}{10}
\emline{46.00}{6.00}{11}{40.67}{4.33}{12}
\bezier{184}(6.67,36.33)(8.00,11.00)(27.67,6.00)
\put(27.00,3.33){\makebox(0,0)[cc]{$R$}}
\put(42.67,10.67){\makebox(0,0)[cc]{$r$}}
\put(12.67,41.00){\makebox(0,0)[cc]{$n(r)$}}
\put(24.33,41.33){\makebox(0,0)[cc]{(a)}}
\emline{66.00}{6.00}{13}{66.00}{44.33}{14}
\emline{66.00}{44.33}{15}{64.00}{38.33}{16}
\emline{66.00}{44.33}{17}{68.00}{38.00}{18}
\emline{66.00}{6.33}{19}{105.00}{6.33}{20}
\emline{105.00}{6.33}{21}{100.33}{8.33}{22}
\emline{105.33}{6.33}{23}{100.33}{4.67}{24}
\put(63.67,19.00){\makebox(0,0)[rc]{$\sqrt{2}$}}
\put(71.00,40.67){\makebox(0,0)[lc]{$d$}}
\put(101.67,12.00){\makebox(0,0)[cb]{$\sigma$}}
\put(84.67,41.00){\makebox(0,0)[cc]{(b)}}
\bezier{164}(66.00,6.33)(65.33,22.33)(90.67,21.33)
\emline{90.67}{21.33}{25}{90.67}{18.67}{26}
\emline{90.67}{16.67}{27}{90.67}{13.67}{28}
\emline{90.67}{11.67}{29}{90.67}{8.33}{30}
\emline{90.67}{6.33}{31}{90.67}{6.33}{32}
\put(91.00,4.00){\makebox(0,0)[ct]{$9/2$}}
\emline{90.67}{21.67}{33}{88.33}{21.67}{34}
\emline{86.00}{21.67}{35}{83.00}{21.67}{36}
\emline{80.00}{21.67}{37}{77.00}{21.67}{38}
\emline{74.67}{21.67}{39}{71.33}{21.67}{40}
\emline{68.33}{21.67}{41}{66.00}{21.67}{42}
\end{picture}
\caption{\scriptsize (a) --- schematic dependence of $n(r)$ for the tips
potential; (b) --- dependency of equilibrium size of the system $d=R/\rho$ in
effective interaction radius units on interaction parameter $\sigma=3\bar
sN/8\pi$.}\label{graf2}
\end{figure}
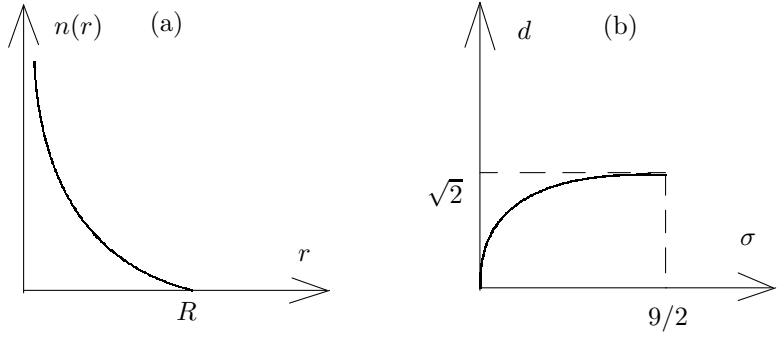

Lets consider a system with potential of the following form:
\begin{equation}\label{yama}
g=\left\{
\begin{array}{rr}
-1,&r<\rho;\\
\\
0,&r>\rho,
\end{array}
\right.
\end{equation}
where $\rho$ --- length parameter, having the sense of effective interaction
radius. The formulae (\ref{kern}) in this case is invalid, and direct
calculation by (\ref{funcg}) gives:
\begin{equation}\label{ykern}
G(r,r')=-\left(\frac{\rho^2-(r-r')^2}{2rr'}\right).
\end{equation}
The integral $K$ for the first approximation (\ref{ex1}) has the form:
\begin{equation}\label{appr}
K=-\frac{1}{2r}\left(\frac{R^2}{2}(\rho^2-\frac{R^2}{2})-\frac{r^2R^2}{2}+
\frac{2rR^3}{3}\right).
\end{equation}
Requirement of vanishing of expression $n=n_0-n_0^2\bar sK$ under $r=R$ gives
$n_0=-4/\bar sR(\rho^2-R^2/6)$. Under $\rho^2=R^2/6$ following approximations
should be accounted. Expression for $n$ after some algebra takes the form:
\begin{equation}\label{concy}
n=-\frac{4}{\bar
sRr(\rho^2-R^2/6)^2}\left(r^2R+r(\rho^2-\frac{3}{2}R^2)-R(\rho^2-R^2/2)\right).
\end{equation}
Its schematic kind is shown in fig.\ref{graf2}(a). Note, that in accordance
with theorem 2, cusp --- is induced by approximation.

Nonnegativity condition for (\ref{concy}) will be nonequality:
\begin{equation}\label{noneq} R\le\sqrt{2}\rho,
\end{equation}
that physically (in this approximation) means coincidence of orders magnitudes
of size of the system and radius of pair interaction. Normalizing condition
(\ref{norm}) then gives:
\begin{equation}\label{ynorm}
\frac{8\pi}{3\bar s}\frac{R^2\rho^2}{(\rho^2-R^2/6)^2}=N.
\end{equation}
With (\ref{noneq})  the (\ref{ynorm}) leads to the following restriction on a
number $N$:
\begin{equation}\label{cond}
\bar sN\le12\pi
\end{equation}
Physical sense  of (\ref{cond}) (which should remain valid, in principle, for
any  approximation) is concluded in the fact, that when interaction constant
$\bar s$ is fixed, there is no equilibrium state for infinitely large $N$,
since adding new particles we don't change attractive interaction intensity,
while repulsive pressure  is increased. This property simulate well known
"saturation" phenomenas of water steam or nuclear systems.

Dependency of equilibrium size  $d=R/\rho$ on the interaction parameter
$\sigma=3\bar sN/8\pi$, $(0\le\sigma<9/2)$ has the kind: $$
d=3\sqrt{2}\sqrt{\frac{1+\frac{\displaystyle\sigma}{\displaystyle
3}-\sqrt{\displaystyle 1+\frac{\displaystyle 2}{\displaystyle
3}\sigma}}{\sigma}} $$ and is shown in fig.\ref{graf2}(b).

\section{Virial correction calculation.}\label{vir}

Lets go backward to the  expression (\ref{eq1})  and lets substitute in it the
more general state equation (\ref{eqst}). Using equality $p'_{n}=a\gamma
n^{\gamma-1}+ab(\gamma+1)n^{\gamma}$, we get:
\begin{equation}\label{eqb}
a\gamma n^{\gamma-1}n'+ab(\gamma+1)n^{\gamma}n'+
2\pi\alpha^2n\frac{\partial\hat K}{\partial r}[n]=0.
\end{equation}
Dividing the both sides over  $n$ and integrating over  $r$ we go to equations
generalizing (\ref{gneq1}) and (\ref{geq1}):
\begin{equation}\label{gneq1b}
\frac{a\gamma}{\gamma-1}n^{\gamma-1}+ab\frac{\gamma+1}{\gamma}n^{\gamma}
+2\pi\alpha^2\hat K[n]+C=0; \ \ (\gamma\neq1);
\end{equation}
\begin{equation}\label{geq1b}
a\ln n+2ab n+2\pi\alpha^2K+C=0,\ \ (\gamma=1).
\end{equation}
We'll find solution to (\ref{gneq1b})-(\ref{geq1b}) by perturbation theory with
small nonlinearity parameter $b$:
\begin{equation}\label{row1}
n(r)=\bar n(r)+b\delta_1 n+b^2\delta_2 n\dots,
\end{equation}
where $\bar n$ --- solution to (\ref{gneq1b})-(\ref{geq1b}) under $b=0$, which
has been obtained earlier, $\delta_1 n, \delta_2 n,\dots$ --- unknown
functions, defining corrections of  different orders. Substituting (\ref{row1})
in  (\ref{gneq1b})-(\ref{geq1b}) and holding terms of a first order over $b$
from the both side we get:
\begin{equation}\label{eqbn}
\delta_1 n+\frac{\gamma+1}{\gamma^2}\bar n^2+\frac{2\pi\alpha^2}{\gamma a \bar
n^{\gamma-2}}\hat K[\delta_1 n]=0,
\end{equation}
--- linear nonhomogeneous  integral Fredholm equation on
$\delta_1 n$. This equation is valid for the case $\gamma=1$ too.

Lets find virial corrections for the considerd earlier potentials in case
$\gamma=1$.

a) {\bf Coulomb potential.} Expression for the kernal  $G$ of  $\hat K$, as it
follows from general formulae (\ref{kern}) has the form:
\begin{equation}\label{kernkul}
G(r,r')= \left\{
\begin{array}{ll}
-2/r,&r'<r,\\ -2/r',&r'>r.
\end{array}
\right.
\end{equation}
Equation (\ref{eqbn}) with the kernal (\ref{kernkul}) and function $\bar n$
from (\ref{conc1}) takes the form:
\begin{equation}\label{eqbnq}
\delta_1 n+\frac{9}{8\bar s^2 R^4}\left(1-\frac{r^2}{R^2}\right)^2+
\frac{3}{4R^2}\left(1-\frac{r^2}{R^2}\right) \left\{
-\frac{2}{r}\int\limits_{0}^{r}r'^2\delta_1n\cdot dr'-2\int\limits_{r}^{R}
r'\delta_1 n\cdot dr'\right\}=0.
\end{equation}
It is convenient to go to the dimensionless variables $z=r/R$, $y=\delta_1
n\cdot R^6$ and to introduce dimensionless parameter $\xi^2=R^2/\bar s^2$. Then
equation (\ref{eqbnq}) will take the following more simple form:
\begin{equation}\label{eqbnq1}
y+\frac{9}{8}\xi^2(1-z^2)^2+\frac{3}{4}(1-z^2) \left\{
-\frac{2}{z}\int\limits_{0}^{z}z'^2y\cdot dz'-2\int\limits_{z}^{1} z'y\cdot
dz'\right\}=0.
\end{equation}
Dividing over $-3(1-z^2)/2z$ and differentiating twice over $z$, we get
differential consequence  of (\ref{eqbnq1}), which is differential equation of
second order:
\begin{equation}\label{diff}
u''+\frac{3}{2}(1-z^2)u+\frac{9}{2}\xi^2z=0,
\end{equation}
where $u=-2zy/3(1-z^2)$. Its solution is given by the power row
\begin{equation}\label{rowb}
u(\xi,z)=\sum\limits_{k=0}^{\infty}a_kz^k,
\end{equation}
where coefficients are defined by the relations:
\begin{equation}\label{coeff}
a_2=-\frac{3}{4}a_0,\ a_3=-\frac{1}{4}(a_1+3\xi^2),\ \
a_{k+2}=\frac{3(a_{k-2}-a_k)}{2(k+1)(k+2)},\ \ k=2,\dots
\end{equation}
Here $a_0$ and $a_1$ --- yet unknown integration constants. To find ones it is
necessary to restrict obtained general solution  to a subclass, satisfying
(\ref{eqbnq1}). Substituting  (\ref{rowb}) in (\ref{eqbnq1}) and integrating
term by term, we equate coefficients for the same powers of $z$. This, apart
from (\ref{coeff}), will give additional equalities: $$ a_0=0,\ \
a_1=\frac{3}{4}\xi^2+3\sum\limits_{n=0}^{\infty} \frac{a_n}{(n+1)(n+3)}, $$
that fully defined our solution. Since $a_0=0$ implies  by (\ref{coeff})
vanishing of all even coefficients in (\ref{rowb}), we finally get:
\begin{equation}\label{pop1}
\delta_1n=-\frac{3(1-z^2)}{2R^6}\sum\limits_{n=0}^{\infty} \beta_nz^{2n},\ \
\mbox{\rm where}
\end{equation}
$$ \beta_0=\frac{3}{4}\left(\xi^2+\sum\limits_{n=0}^{\infty}
\frac{\beta_n}{(n+1)(n+2)}\right),\ \ \beta_1=-\frac{1}{4}(\beta_0+3\xi^2),\ \
\beta_{n}=\frac{3(\beta_{n-2}-\beta_{n-1})}{4n(2n+1)}. $$

From the (\ref{pop1}) after some  analysis we conclude that under $b>0$, (that
takes place for a majority of real systems and means additional repulsion) the
correction  $b\delta_1 n$ is always negative: system become more rared and its
equilibrium size increased, as it is intuitively  obviously.

b) {\bf Harmonic potential}. Taking expression for $\bar n$ from (\ref{conc2})
and substituting it in  (\ref{eqbn}) we get
\begin{equation}\label{eqbnh}
y+2k^2\xi^2(1-z^2)^2+k(1-z^2)\hat K[y]=0,
\end{equation}
where $y$ and $z$ --- are the same, that in previous case, $k=75/128$,
$\xi=1/(\bar sR^2)$. Formulae  (\ref{kern}) for $G$ in case  $\lambda=2$ gives:
\begin{equation}
G(r,r')=2(r^2+r'^2),
\end{equation}
so for $\hat K[y]$ we get
$$ \hat
K[y]=2\int\limits_{0}^{1}z'^2(z^2+z'^2)y(z')dz'= Az^2+B,$$
  where
$A=2\int\limits_{0}^{1}y z'^2dz'$, $B=2\int\limits_{0}^{1}y z'^4dz'$. So the
equation (\ref{eqbnh}) on $y$ is purely algebraic and its solution is reduced
to finding of unknown $A$ and $B$. To do it we substitute $y$ from
(\ref{eqbnh}) in expressions for $A$ and $B$ and after some transformations we
go to the following linear system:
\begin{equation}\label{lin}
\begin{array}{lclcr}
(2kI_{41}+1)A&+&2kI_{21}B&=&-4k^2\xi^2I_{22};\\
2kI_{61}A&+&(2kI_{41}+1)B&=&-4k^2\xi^2I_{42};
\end{array}
\end{equation}
where
\begin{equation}\label{int}
I_{pq}\equiv\int\limits_{0}^{1}x^p(1-x^2)^q\,dx=\sum\limits_{k=0}^{q}
(-1)^{q-k}C^q_k\frac{1}{2(q-k)+p+1}.
\end{equation}
For the integrals in  (\ref{lin}) the (\ref{int}) gives the following
expressions: $$ I_{21}=\frac{2}{15};\ \ I_{22}=\frac{8}{105};\ \
I_{41}=\frac{2}{35};\ \ I_{42}=\frac{8}{315};\ \ I_{61}=\frac{2}{63}. $$
Substituting it into (\ref{lin}) and expressing $A$ and $B$ by Kramers rule, we
get $$ A=-\frac{375}{15808}\xi^2\approx-0.024\xi^2;\ \
B=-\frac{8025}{15808}\xi^2\approx-0.51\xi^2; $$ Finally,
\begin{equation}\label{pop2}
\delta_1 n=-\frac{0.39}{\bar s^4R^{10}}(1-2.8z^2+1.8z^4).
\end{equation}
As it has been in the previous case, the case $b>0$ rares the system.

\section{Conclusion.}

So, the considered approach  is relevant to the following problems:

1) Calculating of equilibrium distribution of any LRIS;

2) Finding of potential parameters by comparing calculated distribution with
observation (inverse to 1-st), that is more important from theoretical point of
view, than 1-st.

3) Calculation of virial corrections.

The following obvious generalizations are possible:

1) generalization on the case of coexisting chemical phases;

2) generalization on rotating systems;

3) generalization on identical quantum particles. This last case will involve
nonlinear self-consistent multiparticle Shredinger equation.

In conclusion I will be grateful to express many thanks to S.B.Moskowskiy for
the interest and useful critical comments.

\appendix
\section{Appendix 1: coefficients
$B^n_{k_1\dots k_n}$.}

Coefficients $B^n_{k_1\dots k_n}$, appearing in the rows (\ref{formula}) can be
calculated for any $n,k_1,\dots k_n$ with the help of the following recurrent
formulae:
\begin{equation}\label{B}
\begin{array}{l}
B^n_{k_1\dots k_{n-1}0}=B^{n-1}_{k_1-1\,k_2\dots k_{n-1}}+
\sum\limits_{j=1}^{n-2}(k_j+1)B^{n-1}_{k_1\dots k_j+1\,k_{j+1}-1\dots
k_{n-1}},\\
\\
B^{n}_{00\dots01}=1.
\end{array}
\end{equation}
All another coefficients are zero, and beginning value $B^{1}_1=1.$ For a
number $f(n)$ of nonzero $B^{n}$ for a given  $n$ the following estimate can be
obtained: $$f(n)<\frac{e^n}{\sqrt{2\pi n}}$$.

Lets give the nonzero $B^{n}_{k_1\dots k_n}$ for $n=\overline{1,5}$
includingly.

\begin{equation}
\begin{array}{llllllll}
n=1& B_1^1=1&\\
\\
n=2& B^2_{20}=1;   & B^2_{01}=1;\\
\\
n=3& B^3_{300}=1;  & B^3_{110}=3;     &  B^3_{001}=1;\\
\\
n=4& B^4_{4000}=1; & B^4_{2100}=6;    & B^4_{1010}=4;  & B^4_{0200}=3; &
B^4_{0001}=1;\\
\\
n=5& B^5_{50000}=1;& B^{5}_{31000}=10;& B^5_{20100}=10;& B^5_{12000}=15;&
B^5_{10010}=5;\\
\\
& B^5_{01100}=10;& B^5_{00001}=1
\end{array}
\end{equation}

\section{Appendix 2: calculation of  $K_n$}

Lets show the results of calculations of integrals, which are necessary for
third correction calculation.

For a Coulomb potential: $$
\begin{array}{l}
\displaystyle K_1=\frac{r^2}{3}-R^2;\\
\\
\displaystyle K_2=\frac{r^4}{30}-\frac{r^2R^2}{3}+\frac{5R^4}{6};\\
\\
\displaystyle
K_3=\frac{r^6}{630}-\frac{R^2r^4}{30}+\frac{5r^2R^4}{18}-\frac{61R^6}{90};\\
\\
\displaystyle
K[K^2]=\frac{r^6}{189}-\frac{R^2r^4}{15}+\frac{r^2R^4}{3}+\frac{19R^6}{27}.
\end{array}
$$

For harmonic potential:
$$
\begin{array}{l}
\displaystyle K_1=\frac{2r^2R^3}{3}+\frac{2R^5}{5};\\
\\
\displaystyle K_2=\frac{8R^8r^2}{15}+\frac{184R^{10}}{525};\\
\\
\displaystyle K_3=\frac{704R^{13}r^2}{1575}+\frac{768R^{15}}{2625};\\
\\
\displaystyle K[K^2]=\frac{704R^{13}r^2}{1575}+\frac{22336R^{15}}{70875}.
\end{array}
$$ 

\end{document}